\documentclass[manuscript]{acmart}

\AtBeginDocument{%
  }

\setcopyright{acmlicensed}
\copyrightyear{2026}
\acmYear{2026}
\acmDOI{https://doi.org/10.1145/3821700}

\acmConference[Conference acronym MobileHCI'26]{Make sure to enter the correct
  conference title from your rights confirmation emai}{August,
  2026}{Swansee, Wales, UK}

\acmISBN{978-1-4503-XXXX-X/18/06}
\usepackage{soul}
\usepackage[normalem]{ulem}

\usepackage{booktabs}
\usepackage{multirow}
\usepackage{tabularx}

\begin{document}

\setcopyright{cc}
\setcctype{by}
\acmJournal{PACMHCI}
\acmYear{2026} \acmVolume{10} \acmNumber{5} \acmArticle{MHCI8950}
\acmMonth{8} \acmDOI{10.1145/3821700}

\title{“When Suspicion Becomes Detection”: Folk Deception Cues and Detection Strategies in Online Dating Romance Scams}






\author{Sima Amirkhani}
\orcid{0009-0009-8836-8291}
\email{Sima.Amirkhani@h-brs.de}
\affiliation{%
\institution{University of Siegen}
\city{Siegen}
\country{Germany}
}

\author{Jana Krüger }
\orcid{0000-0003-3117-6802}
\affiliation{%
\institution{Independent Researcher}
\country{Germany}}

\author{Dave Randall}
 \orcid{0000-0001-8613-3477}
\affiliation{%
\institution{University of Siegen}
\city{Siegen}
\country{Germany}}

\author{Gunnar Stevens}
 \orcid{0000-0002-7785-5061}
\affiliation{%
\institution{University of Applied Science Bonn-Rhein-Sieg}
\city{Sankt Augustin}
\country{Germany}}

\author{Douglas Zytko}
\orcid{0000-0002-6854-5336}
\affiliation{%
\institution{University of Michigan-Flint}
\city{Flint}
\state{Michigan}
\country{United States}
}

\renewcommand{\shortauthors}{Amirkhani et al.}

\begin{abstract}
The growth of mobile dating platforms has coincided with a rise in romance scams, in which offenders construct convincing personas to defraud users. While research on romance scams is expanding, victims’ lived experiences of recognizing and responding to deception in mobile-mediated interactions remain insufficiently understood. To address this gap, we conducted in-depth interviews with 24 victims of online dating romance scams in Iran, where legal, social, and cultural constraints limit formal support. Our analysis identifies suspicion cues and the investigative strategies victims use to verify identities across platforms. We show that victims are not passive recipients of deception but engage in active, iterative detection practices under significant emotional, social, and relational pressure. Based on these findings, we contribute empirically grounded insights into deception cues and user-driven detection work, and we discuss implications for the design of mobile technologies that better support users in identifying, resisting, and recovering from romance scams.
\textit{\textbf{Content Warning:} This paper discusses sexual violence.}

\end{abstract}

\ccsdesc[500]{Security and privacy~Social aspects of security and privacy}

\keywords{Online Dating Romance Scam, Deception cues, Detection Strategy, Iran}


\maketitle

\section{Introduction}


The widespread adoption of mobile dating platforms 
as transformed how intimate relationships are initiated and maintained, but it has also enabled new forms of interpersonal deception. Online dating romance scams involve the strategic use of fabricated or misrepresented identities to cultivate emotional intimacy and trust, with the goal of extracting resources such as money, gifts, or sexual labor \cite{whitty2013scammers, garrett2014exploring, amirkhani2024beyond, feng2026grooming}. By exploiting the affective dynamics of romantic relationships, offenders leverage intimacy itself as a mechanism of fraud \cite{ANESA}. Beyond financial loss, these scams frequently result in severe emotional, psychological, and social harm, including shame, distress, and long-term erosion of interpersonal trust \cite{lazarus2023we, wang2026scoping}.

Although the scale of online dating romance scams is substantial \cite{cross2022suspect}, victimization is widely under-reported. One key reason for this under-reporting is that many victims do not initially recognize their experiences as fraud \cite{button2017cyber}. When deceptive interactions are interpreted as genuine relationships—or as personal misjudgments rather than criminal acts—individuals are unlikely to seek help or report harm. As a result, the processes through which users come to recognize romance scams, including how suspicion emerges and how users make sense of ambiguous or contradictory cues, remain insufficiently understood. This gap constrains the development of effective prevention strategies and user-centered safety interventions. 

Research on deception detection has a long interdisciplinary history and has intensified in recent years \cite{bullee2018anatomy}. Prior work has identified a range of verbal, behavioral, and relational cues associated with deceptive conduct \cite{depaulo2003cues, veisi2025user}. Within online dating contexts, such cues can function as early warning signs that prompt users to reassess the authenticity of an interaction. However, much of the existing literature on romance scams has been conducted in Western contexts. 

Romance fraud is a global phenomenon, with rising victimization documented across the world \cite{cross2023knew}. While Western contexts have received the bulk of scholarly and institutional attention, the impacts of romance scams can be more acute in non-Western settings, where victims face heightened stigma, legal risk, or limited access to formal support due to unique local sociocultural factors \cite{amirkhani2026my, abubakari2024espouse, johnson2025digital, ghadamighalandari2025between}.  
For instance, in some cultures becoming the victim of an online dating romance scam can justify being murdered by one's father because of the familial shame incurred by the scam's discovery \cite{amirkhani2025society}. Despite these unique vulnerabilities, and need for preventative solutions given the mortal (and not just financial) risks of scam victimization, non-Western contexts remain comparatively underrepresented in research about deception detection during online dating romance scams. This under-representation is particularly consequential in settings where romantic relationships are legally restricted or socially stigmatized, as such conditions may fundamentally alter how deception is experienced, recognized, and disclosed.





We address this gap with a study of 
 24 in-depth narrative interviews with Iranian victims of online dating romance scams. Iran offers a particularly revealing case, as premarital romantic relationships are legally prohibited and socially stigmatized. Individuals who pursue such relationships often do so covertly, navigating significant moral, familial, and legal risks \cite{cucci2019adolescent, muniz2019parental, amirkhani2024beyond}. In this context, mobile dating applications and private messaging platforms have become critical spaces for forming intimate connections, especially among younger users \cite{emami2021girlboy, amirkhani2024beyond}. At the same time, these conditions may intensify vulnerability to deception while constraining victims’ ability to act on, or seek support regarding, suspicions of deception.
Our research questions were:
\begin{itemize}
\item \textbf{\textit{RQ1: What behavioural and interactional cues do Iranian victims retrospectively identify as prompting suspicion in online dating romance scams?} }
\item \textbf{\textit{RQ2: How do Iranian victims interpret these suspicions, and what investigative actions do they describe taking in response?}}
\end{itemize}

While in some prior work victims have recognized romance fraud through distinct behavioral "triggers" \cite{cross2023knew}, the literature generally finds that scam detection or realization is an iterative and gradual process \cite{cross2022suspect, whitty2013scammers, vrij2015deception, jong2019detecting}. We similarly find that scam suspicion-turned-detection is a gradual process, albeit one uniquely shaped by local cultural factors. In our non-Western cultural context of Iran, local norms against premarital relationships and expectations for early marriage, as well as legal villainization of romance scam \textit{victims} rather than perpetrators, imposed barriers on investigative processes by victims that stifled the pace of scam confirmation and actively imposed doubt on victims' own suspicions. In light of traditional dating apps being banned in Iran, these investigative practices were further hindered by the distributed network of platforms used by scammers for contacting and engaging with their victims. The same cultural dynamics that impede victims' responses to scam suspicion also limit the applicability of Western-centric solutions for scam detection, such as platform-specific features (given that dating apps are banned in Iran) and legal or law enforcement-driven approaches (given that scam victimization itself is typically considered a crime in our study's context).

This paper makes three contributions. \textit{First}, it expands the geographic and cultural scope of romance scam research by providing an empirically grounded account of how online dating romance scams are retrospectively recognized and made sense of in a non-Western context by victims navigating emotionally invested and culturally-specific interactions. \textit{Second}, it foregrounds victims’ culturally-situated investigative labor, highlighting how users actively assess authenticity under conditions of uncertainty that are amplified  by cultural stigma against romance scams and prematerial relationships, as well as constraints on publicly discussing potential victimization. \textit{Third}, it argues for culturally-sensitive design interventions when traditional dating apps are not used for online dating scams and typical legal recourse is not available to victims. Specifically, we make a case for third party whisper network-oriented approaches to augmenting romance scam detection in non-Western contexts.

\section{Related Work}
Online dating romance scams are a form of interpersonal scam in which offenders create fabricated or strategically misrepresented identities on dating applications or social networking platforms to deceive victims under the guise of a romantic relationship \cite{whitty2015anatomy, buchanan2014online, whitty2012better}. These scams fundamentally rely on deception \cite{jimoh2018love, cross2020romance}, often involving stolen photographs, false biographical details, and carefully curated personas—frequently drawing on socially trusted archetypes such as military personnel \cite{cross2022suspect}. As Whitty argues, romance scams are defined by pretense: the deliberate construction of an intimate relationship that does not exist \cite{whitty2015anatomy}. 

From the scammer's perspective, prior research has documented a relatively consistent pattern in scammer behavior. Typically, scammers initiate frequent and emotionally intensive communication to establish intimacy and trust \cite{whitty2013scammers, benedict2025love}. Over time, they introduce obstacles that prevent in-person meetings and frame requests for money or other resources as necessary to overcome these barriers \cite{garrett2014exploring}. Even when initial requests are fulfilled, scammers often continue the deception to extract additional value. Despite the scale and harm associated with online dating romance scams, victimization remains substantially under-reported, in part because victims may not initially recognize their experiences as fraud. 

From the scam victim's behavior, prior work in fraud victimology, deception detection, and social engineering has shown that scam recognition often unfolds incrementally, in part due to victims initially rationalizing early warning signs through emotional and cognitive biases \cite{alizadeh2023catch, cross2016they, kopp2015role, lacey2015taking, pins2024digitaler, whitty2012online}. 

The following sections provide a more in-depth review of prior online dating romance research on (1) red flags and deception cues and challenges victims face in recognizing romance scams in section \ref{Red Flags}, (2) existing approaches to scam detection and investigation in section \ref{detection}, and (3) the legal and social context of dating in Iran in section \ref{Iran}. (4) We conclude by synthesizing gaps that motivate our study in section \ref{overlooked}.

\subsection{Red Flags and Scam Suspicion Cues} \label{Red Flags}
A “red flag” refers to conditions or patterns that deviate from expectations and may warrant further investigation \cite{dinapoli2008red, baader2018reducing}. Red flag approaches are widely used in fraud auditing and prevention, where they serve as heuristic indicators of potential misconduct rather than definitive proof \cite{baader2018reducing}. As DiNapoli notes, fraud typically escalates gradually, becoming noticeable only once inconsistencies accumulate or deviations become salient \cite{dinapoli2008red}. In early stages, fraudulent behavior may be indistinguishable from legitimate interaction.
Importantly, red flags do not establish guilt or deception on their own. Rather, they function as signals that raise doubt and invite further scrutiny \cite{baader2018reducing}. An isolated red flag may be easily rationalized, whereas multiple cues, when interpreted together, can prompt suspicion. This moment of suspicion is critical, as it often precedes investigative action and eventual scam detection \cite{marimuthu2022intelligent}. However, prior research has largely treated red flags as static indicators, giving limited attention to how users interpret, negotiate, or sometimes dismiss these cues within ongoing romantic interactions.

\subsubsection{Deception Cues in Online Dating Romance Scam} \label{Deception Cues}
Deception involves the intentional attempt to create a false belief or sense of trust in others \cite{vrij2015deception, barkallah2026wanted}. Within online dating contexts, deception operates along a spectrum, ranging from minor exaggerations to elaborate identity fabrication \cite{hancock2007truth, marimuthu2022intelligent}. Research suggests that deceptive individuals are often less forthcoming, provide less coherent narratives, and exhibit higher levels of inconsistency and emotional manipulation than truth-tellers \cite{depaulo2003cues}. Narratives that appear “too good to be true” have also been associated with deceptive intent \cite{depaulo2003cues, harrington2009deception}.

Studies of online dating romance scammers further demonstrate that adopting an appealing persona and masking one’s true identity is a deliberate and strategic tactic \cite{cross2022suspect, dickinson2023money}. Scammers frequently mirror victims’ desired partner traits, reinforce emotional dependency, and manipulate trust to elicit compliance \cite{dickinson2023money}. Other commonly identified deception cues include the use of multiple accounts \cite{tsikerdekis2015detecting}, inconsistencies in personal information \cite{jimoh2018love}, avoidance of in-person meetings, and explicit requests for money or gifts \cite{cross2024romance}. Scammers may also employ relational manipulation strategies such as ghosting, breadcrumbing, and emotional withholding, which keep victims in a prolonged state of uncertainty and anticipation \cite{navarro2020ghosting, khattar2023young, lee2007judging, cross2023knew}.

While this body of work has been instrumental in identifying common deception cues, it largely focuses on cues as observable indicators identified by researchers, auditors, or third parties—often retrospectively. Less is known about how users themselves experience, interpret, and act upon these cues as part of an unfolding, emotionally invested interaction, particularly within mobile dating environments- e.g., dating through social media.

\subsubsection{Why Victims Struggle to Detect Online Dating Romance Scams}
Prior research in fraud victimology, deception detection, and social engineering has already demonstrated that scam engagement and detection are often iterative, gradual, and psychologically mediated processes rather than discrete decision points (e.g., \cite{lacey2015taking, lacey2024cyberpsychology, kopp2015role, kopp2016online, cross2016they, rege2009s}). For instance, Kopp \cite{kopp2016online} conceptualizes romance scams as phase-based relational processes that progressively build trust and shape victim responses, while Lacey \cite{lacey2024cyberpsychology} highlights the role of cognitive, emotional, and social factors in shaping compliance across scam contexts.

Detecting deception in romantic contexts is particularly challenging due to the emotional intensity of intimate relationships. Prior research shows that victims are less likely to suspect fraud when emotional investment is high \cite{whitty2012online, cross2021use}. Norms associated with romantic love emphasize trust, openness \cite{amirkhani2026talking, amirkhani2026let}, and emotional vulnerability, making suspicion feel psychologically or morally inappropriate \cite{atkins2021epistemic, sorell2019online, amirkhani2025privacy}. As Lee et al. \cite{lee2023self} demonstrate, scammers exploit these norms by crafting profiles and messages that foreground commitment, affection, and long-term relational intent. Individual characteristics further shape vulnerability. People with strong romantic ideals or high emotional openness have been found to be more susceptible to online dating romance scams \cite{whitty2012better}. Scammers may reinforce this susceptibility by constructing a shared history—invoking shared memories, secrets, or future plans—which anchors victims emotionally and discourages critical scrutiny \cite{whitty2019can}. 

Detection is further complicated in online environments, where the absence of physical, auditory, and contextual cues limits users’ ability to verify authenticity \cite{krombholz2015advanced}. More broadly, research suggests that humans are generally poor at detecting deception, even when cues are present \cite{rege2009s, whitty2015anatomy}. Notably, recognizing deception does not always lead to disengagement. Some victims continue interacting despite suspicion, viewing the situation as a calculated risk or hoping the relationship may still be genuine \cite{lea2009psychology}. Others may struggle to disengage even when they believe deception is likely \cite{sorell2019online}. 

So, this literature suggests that detecting deception in romantic contexts is not a simple matter of noticing red flags, but an ongoing interpretive process shaped by emotional attachment, relational norms, and situational constraints. Suspicion may emerge gradually, fluctuate over time, or coexist with hope and doubt, rather than leading directly to disengagement. Yet, existing research has largely treated detection as an outcome—whether victims “noticed” the scam or not—rather than examining how individuals actively reason through uncertainty, test competing interpretations, and weigh the social and emotional costs of acting on suspicion. This gap is particularly pronounced in contexts where disclosure, confrontation, or disengagement carries heightened moral, social, or legal risk.

\subsection{Detection and Fraud Investigation} \label{detection}
Prior research has proposed several strategies for detecting online dating romance scams, often framed as best practices or preventative recommendations. One commonly suggested approach is the “authenticity check,” such as arranging in-person meetings early in the relationship to reduce the likelihood of deception \cite{jimoh2018love}. Technological solutions, including biometric authentication or identity verification, have also been proposed as potential safeguards \cite{jimoh2018love}.
Internet-based investigation is another frequently cited strategy. Searching profile photos, names, or biographical details across platforms can reveal inconsistencies or reused scam content \cite{cross2022suspect}. Cross and Layt’s analysis of victim reports highlights internet searches and consulting family or friends as common detection strategies \cite{cross2022suspect}. Social connections within platforms may also help expose fraudulent identities \cite{jimoh2018love, ankee2015review}.

However, much of this literature treats detection as a rational, low-cost activity and offers prescriptive advice rather than empirically grounded accounts of user behavior. These strategies may be difficult to enact in contexts characterized by secrecy, stigma, or legal risk, and they provide limited insight into how users navigate suspicion in practice—particularly within emotionally charged, mobile-mediated interactions. Collectively, this literature conceptualizes fraud detection as a rational and relatively low-cost set of actions, leaving underexplored how users experience suspicion, negotiate risk, and decide whether to act under emotional, social, or legal constraints.

\subsection{Dating Norms and Constraints in Iran} \label{Iran}
\textit{Content warning:} This section discusses honor-based violence and legal sanctions.

Because this study focuses on Iranian online daters, it is important to briefly outline the legal and social context in which romantic relationships may occur. Prior scholarship suggests that premarital romantic relationships in Iran are legally prohibited and (can be) socially stigmatized, reflecting a broader framework of religious, legal, and cultural norms reinforced after the 1979 revolution \cite{daniel2012history, shaheed2018outlier, khalajabadi2015perceived, shahidian2002gender}. Legal reforms that had previously reflected aspects of Western family law were rolled back following the revolution, accompanied by the reestablishment of more conservative social and legal norms \cite{shahidian2002gender, paidar1997women}. The legal minimum age for marriage was reduced to 9 for girls and 15  for boys \cite{asghari2019early, amirkhani2025society}, while laws permitting polygamy were reinstated and male authority in divorce proceedings was further strengthened \cite{afshar1987women}. Physical or verbal intimacy between unmarried men and women may be criminalized under laws addressing offenses such as \textit{Mozajeeh}, \textit{Taqbil}, and \textit{Zena}, with penalties ranging from corporal punishment to severe legal sanctions \cite{boostani2022iranian, rahbari2022marriage}.

Prior scholarship and human rights reporting have documented that, in some conservative familial and regional contexts in Iran, perceived violation of norms surrounding female sexuality, premarital relationships, or family honor may contribute to risk of honor-based violence against women and girls \cite{schneider2012concept, pirnia2020honour}. As a result, individuals—particularly younger generations—often engage in premarital relationships covertly to avoid familial, social, and legal repercussions \cite{farahani2015meta}. In this context, online and mobile social media offer a critical space for initiating and maintaining romantic relationships with greater perceived privacy and control \cite{farahani2015meta, emami2021girlboy}. At the same time, these conditions may amplify vulnerability to deception while constraining victims’ ability to seek advice, disclose harm, or access formal support.

\subsection{Synthesis and the Overlooked Intersections}\label{overlooked}
Prior work examining how users investigate romance scams has primarily relied on secondary data sources. Notably, Cross and Layt \cite{cross2022suspect} and Cross \cite{cross2022suspect}—to our knowledge the only studies that explicitly examine victim investigative strategies—base their analyses on textual complaints submitted to the Australian Scamwatch portal. While such data offers valuable large-scale insights, it captures post hoc, institutionally framed accounts of victim experiences. This methodological choice may obscure the situated, emotional, and interactional processes through which suspicion emerges and investigative actions are undertaken.



Furthermore, in a scoping review of online dating romance scams, Wang \cite{wang2026scoping} shows that existing research remains heavily concentrated in Western contexts, with limited attention to linguistic, cultural, and regional diversity beyond a small number of settings, such as Southeast Asia. The review calls for studies that examine how sociocultural factors—including gender norms, economic pressures, and cultural attitudes toward intimacy and finance—shape both victim vulnerability and offender strategies, emphasizing the need for culturally grounded, empirically situated research. Building on this body of work, 
we seek to empirically examine how such romance scam suspicion and detection unfolds in practice within Iran's unique sociocultural, legal, and interactional settings.

Across existing scholarship, significant progress has been made in identifying romance scam structures, deception cues, and recommended detection strategies. 
However, much of this work either models scam progression at a structural level (e.g., phase-based accounts) or examines detection retrospectively or prescriptively, offering limited fine-grained, in-situ accounts of how users experience, interpret, and act on suspicion during ongoing interactions. In particular, we lack empirical accounts of how victims engage in investigative practices within ongoing, emotionally invested, mobile-mediated interactions—especially in contexts where social stigma and legal constraints heighten risk and limit access to external support. This study addresses this gap by 
providing an empirically grounded account of how victims iteratively recognize, negotiate, and act upon suspicion in practice within live interactions, foregrounding user-led detection work within the socio-cultural context of Iran. 

\section{METHODOLOGY}

To explore our research questions we adopted a qualitative, interview-based approach grounded in first-hand accounts from victims. This enables us to examine user-led investigative practices as they are experienced and narrated in context, rather than reconstructed through institutional reporting channels and other secondary data sources as customary in prior work. Because all participants were based in Iran, where face-to-face interviews posed safety and logistical risks, interviews were conducted remotely via social media platforms. Our analysis focuses on how initial suspicions were raised, what deception cues participants identified, and how they attempted to investigate these suspicions in practice situated within a non-Western context.

\subsection{ Recruitment Strategy and Inclusion Criteria}
Participants were recruited primarily through public posts on an Instagram account administered by the first author, which had approximately 12,000 followers at the time of recruitment. This account was not used for personal communication, but instead functioned as a public-facing space centered on romantic expression in Iran, featuring Persian-language materials such as poetry, short texts, music, and reflections on love and breakups. Although this recruitment strategy may skew participation toward individuals already willing to engage with discussions of romance, it must be understood within the constraints of the Iranian context.

In Iran, formal mechanisms for recruiting research participants—particularly for studies involving sexuality, intimacy, or dating—are extremely limited due to legal, political, and cultural restrictions \cite{rahmani2015sexuality}. Discussions of romance and sexual relationships are widely stigmatized, especially for young people, and institutional recruitment channels are largely inaccessible. Social media spaces oriented around romantic content therefore represent one of the few viable environments in which such topics can be discussed and where affected individuals can be reached \cite{ruitenburg2026we}. Consequently, this recruitment approach was shaped primarily by structural necessity rather than convenience, reflecting the challenges of engaging a hard-to-reach population.

Participation was limited to individuals who self-identified as having experienced deception by a person using a false or misleading identity in an online dating context, resulting in the loss of something of value. Several individuals contacted the research team describing emotional distress following the end of an online relationship but without evidence of deception; these cases were excluded from the study.

Participants reported losses that fell into three broad categories: financial or material loss, sexual exploitation (including coerced sexual acts or content), and emotional investment. These categories correspond to established typologies of romance scam victimization, including Whitty’s persuasive model \cite{whitty2013scammers} and Amirkhani’s extension addressing sexual exploitation (“body scams”) \cite{amirkhani2024beyond}. An overview of participant demographics is provided in Table~\ref{tab:commands}.

\subsection{Positionality Statement} 
The first author is an Iranian woman residing outside Iran and served as the lead interviewer. She maintains strong ties to Iranian online communities through social media and has long engaged with romance-related content as part of public-facing educational and advocacy work. Her visibility and familiarity within these communities facilitated trust-building during recruitment and interviews, particularly given the sensitivity of the topic.

At the same time, this positionality shaped the research process. Participants may have perceived the interviewer as a culturally knowledgeable and sympathetic insider, potentially enabling richer disclosures while also requiring reflexive awareness of shared assumptions. To mitigate unexamined bias, data interpretation was conducted collaboratively among authors with non-Iranian backgrounds, allowing insider cultural knowledge to be balanced with external critical perspectives.

\subsection{Interview Procedure}
All interviews were conducted remotely using social media platforms commonly used in Iran, including Instagram, Telegram, and WhatsApp. Participants were free to choose between text-based and voice-based interviews depending on their preferences for privacy, safety, and comfort. This flexibility was particularly important given the sensitive nature of the topic and participants’ concerns about surveillance, family monitoring, or unintended disclosure.

Interview duration ranged from approximately 40 minutes to 2.5 hours, with longer sessions typically reflecting more prolonged or complex scam experiences. All interviews were conducted in Persian to ensure linguistic clarity and cultural sensitivity. The lead author transcribed the interviews verbatim and subsequently translated them into English, prioritizing consistency and fidelity to participants’ original meanings.

Following each interview, participants were briefly debriefed using open-ended check-in questions (e.g., “How are you feeling right now?”) to assess immediate emotional well-being. While no participant reported acute distress during interviews, three individuals later contacted the researcher seeking professional psychological support related to their scam experiences. These participants were provided with information about vetted Iranian mental health professionals offering online services. In addition, the first author organized two publicly accessible livestreamed discussions with a psychologist focusing on prevention and coping strategies related to romance scams; attendance at these sessions was open to the public and explicitly not linked to study participation.



\subsection{Ethical Considerations}

Ethical approval was obtained from the lead author’s university ethics board. Given the legal and social risks associated with premarital relationships in Iran, extensive measures were taken to protect participants \cite{noh2026protocol}. Participants were assigned pseudonyms, and identifying details such as specific locations were generalized. All data were stored on offline devices, and participants were encouraged to delete chat histories if desired. Interviews were scheduled flexibly, including late-night sessions, and paused or rescheduled if participants expressed concern about being overheard. Communication was designed to be discreet, using neutral greetings and minimizing traceable interaction. Interviews were conducted using a trauma-sensitive approach, with participants free to pause or terminate sessions at any time.

\subsection{ Data Analysis Approach}

We analyzed the interview data using Reflexive Thematic Analysis (RTA) \cite{braun2019thematic, braun2021thematic}, an interpretive qualitative approach that emphasizes meaning-making, researcher reflexivity, and iterative engagement with data. RTA was well suited to this study as it supports the analysis of emotionally 
complex, retrospective narratives while attending and enables attention to sociocultural context and relational dynamics
Rather than treating participants' accounts as objective records of events, we focused on how they constructed and interpreted experiences of suspicion, deception, vulnerability, and investigation.

Analysis followed the phases outlined by Braun and Clarke \cite{braun2021thematic}. First, the lead author conducted data familiarization by transcribing the interviews verbatim in Persian and translating them into English, while repeatedly reading the transcripts to gain an in-depth understanding of participants’ experiences. Second, initial codes were generated inductively, capturing patterns related to suspicion, deception cues, emotional responses, and investigative actions. Coding remained flexible and evolving rather than fixed.

Theme development occurred through iterative cycles of collaborative discussion among four authors. Rather than seeking consensus through intercoder reliability, we adopted a reflexive collaborative approach \cite{braun2021thematic}, treating differences in interpretation as analytic resources. The author with Iranian background provided culturally grounded interpretations of participants’ accounts, particularly regarding family dynamics, stigma, while non-Iranian authors challenged assumptions and prompted elaboration of themes. This collaborative process helped make researcher assumptions visible rather than eliminating them, consistent with RTA's view of researcher subjectivity as an unavoidable and productive part of qualitatiove knowledge production.
Through repeated refinement, codes were organized into broader themes that captured both individual experiences and the sociocultural conditions shaping suspicion and detection practices. Themes were reviewed and refined to ensure coherence, distinctiveness, and analytic depth. The final thematic structure was consolidated during the writing of the Findings section, reflecting the iterative and interpretive nature of RTA -codes evolved iteratively rather than through a fixed codebook.

Nevertheless, to acknowledge this limitation, as a qualitative study based on a relatively small and context-specific sample, our findings are not intended to be statistically generalizable to all online dating romance scams. Instead, our study aims to provide in-depth, contextually situated insights into how suspicion and investigation are experienced and negotiated by our Iranian participants.





\section{Findings }

\begin{table}[t]
\centering
\small
\begin{tabularx}{\textwidth}{p{2.6cm} p{3.6cm} X}
\toprule
\textbf{Main Phase} & \textbf{Subsection} & \textbf{Analytic Description} \\
\midrule

\textbf{1. Suspicion} 
& \textbf{1.1 Suspicion Cues} 
& Accumulation of behavioral and interactional irregularities (e.g., exaggeration, ghosting, sexual requests, avoidance of authentication, social isolation, violations of epistemic norms of love). Cues were often ambiguous in the moment and became legible retrospectively. \\

& \textbf{1.2 Provisional Red Flags} 
& Early warning signs were acknowledged but treated as temporary or explainable. Participants reframed risk as romance, trust as moral obligation, and secrecy as necessity under socio-cultural constraints. \\

& \textbf{1.3 Recognition Without Exit} 
& Participants recognized deception or harm but did not disengage due to emotional dependence, fear of loss, the “gamble” of marriage, and structural vulnerability (stigma, reputational and legal risk). \\

\midrule

\textbf{2. Detection} 
& \textbf{2.1 Searching Across Contexts} 
& Cross-platform and offline searches (names, photos, phone numbers, tagged accounts, acquaintances) to triangulate identity and narrative consistency when direct verification was unavailable. \\

& \textbf{2.2 Questioning and Testing} 
& Indirect probing, repeated verification requests, and consistency tests (including fabricated scenarios or alternate accounts), with direct confrontation typically delayed. \\

& \textbf{2.3 Requests for Proof} 
& Demands for IDs, tickets, selfies, video calls, or meetings—often met with evasions or fabricated evidence that temporarily restored plausibility. \\

& \textbf{2.4 Role of Third Parties} 
& Friends, family members, teachers, therapists, or professionals acted as epistemic intermediaries, validating concerns, investigating independently, enabling disengagement, and supporting recovery. \\

\bottomrule
\end{tabularx}
\caption{Suspicion and detection as an iterative process in online dating romance scams: from emerging cues and deferred doubt to selective investigative action under emotional and socio-cultural constraint.}
\label{tab:suspicion-detection}
\end{table}

This section presents an analytic account of how Iranian victims of online dating romance scams recognized deception cues and responded to emerging suspicions. Rather than treating detection as a single moment of realization, participants’ experiences reveal a gradual, iterative process shaped by emotional investment, social media interaction, and sociocultural constraints. Suspicion rarely emerged from a single incident; instead, it developed through the accumulation of irregularities that participants actively interpreted, negotiated, and sometimes deliberately ignored.
Across accounts, we observed a recurring trajectory: (1) the emergence of suspicion cues, see section \ref{4.1}, (2) the suppression or rationalization of doubt, see section \ref{4.2}, (3) moments of heightened uncertainty, see section \ref{4.3}, and (4) selective investigative action, see section \ref{4.4}. These stages were neither linear nor universal, but they offer an analytic lens for understanding how detection unfolds in practice (as a summary of findings, see table \ref{tab:suspicion-detection}).



\subsection{Suspicion Cues: What Triggered Suspicion}\label{4.1}
Recognizing deception in online dating romance scams was rarely immediate or straightforward. Most participants initially experienced their relationships as authentic and emotionally meaningful, only later recognizing that they had overlooked or rationalized early warning signs. Rather than emerging from a single moment or incident, suspicion typically developed through the gradual accumulation of behavioral irregularities that conflicted with participants’ expectations of romantic interaction.
We conceptualize these irregularities as suspicion cues: patterns of behavior that deviated from participants’ normative understandings of intimacy, commitment, and trust. Importantly, these cues were not always recognized as deceptive at the time they occurred. Although there is some overlap around the meaning of these cues, we present them as distinct categories for analytic clarity; the themes having emerged from how participants themselves interpreted and grouped suspicious behaviors during retrospective sense-making. Participants did not treat all suspicion cues as equivalent; rather, different behaviors became meaningful through different interpretive logics. As an example around differences between "Avoiding Authentication” and “Avoiding Social Connections”: While both themes concerned difficulties in establishing trust, participants distinguished between failed direct verification and blocked external corroboration. "Avoiding Authentication" referred to perpetrators resisting immediate or embodied forms of identity verification (e.g., selfies, instant calls, video calls, or meetings), whereas "Avoiding Social Connections" referred to restricting victims’ access to social networks, mutual contacts, or online visibility that could enable third-party confirmation or accountability.




\subsubsection{"Exaggeration: Intensified Intimacy and Implausible Self-Presentation"}
A prominent suspicion cue reported by seven participants was \textbf{\textit{exaggeration}}, characterized by emotional intensity, accelerated commitment, and grandiose self-presentation that escalated unusually quickly. Victims described perpetrators expressing profound affection, future plans, and promises of marriage within weeks of initial contact. While these behaviors initially felt flattering or romantic, their speed and intensity later became difficult to reconcile with participants’ expectations of relational development.
Betty captured this tension clearly:

\begin{quote}
\textit{“The way he was talking—it was really dreamy! [...] But whenever he presented himself as this incredible person, I became suspicious!”} (Betty)

\end{quote}


What made \textbf{\textit{exaggeration}} salient as a cue was not merely emotional expressiveness, but its disproportionate scale and timing. Several participants described how proposals or declarations of lifelong commitment occurred before meaningful mutual knowledge had formed. Gilda reflected on this temporal mismatch:
\begin{quote}
    \textit{"He wanted to come to propose me. It was unbelievable for me. We just had one and half month together. It was so soon for this request!” (Gilda)}
\end{quote}


\textbf{\textit{Exaggeration}} also appeared in scammers’ self-descriptions. Perpetrators frequently claimed wealth, professional success, or high social status, yet failed to substantiate these claims through everyday actions. Shadi’s account highlights how contradictions between grand narratives and mundane behavior generated suspicion: 
\begin{quote}
   \textit{"Then he said his phone was broken and that he only had a basic phone without a camera. When I questioned why he didn’t just buy a new one—given that he claimed to be rich—he said he had several business loans and couldn’t afford it at that moment. Whenever I asked for a recent photo, he would send me just one picture per month. It felt off." (Shadi)} 
\end{quote}

In some cases, \textit{\textbf{exaggeration}} manifested as emotional pressure. Once affection was established, perpetrators intensified demands by framing compliance as proof of love. Rahil described how this escalation triggered discomfort: 
\begin{quote}
    \textit{“Once he felt sure of my love, he asked for naked photos. He pressured me emotionally, saying that if I truly wanted him, I had to prove it this way. Otherwise, he would end things. But love shouldn’t work like that.”} (Rahil)
\end{quote}

Indeed, these accounts show how \textit{\textbf{exaggeration}} functioned as both an attraction strategy and a source of suspicion—simultaneously accelerating emotional attachment while, within a high-risk relational context, planting early doubts about authenticity and intent.





\subsubsection{"Ghosting: Disappearance as a Signal of Deception"}

Another major suspicion cue involved \textbf{\textit{ghosting}}, characterized by the abrupt withdrawal of communication or an alternation between prolonged absence and minimal contact. Ghosting functioned both as an interactional irregularity and, in some cases, as part of broader patterns of identity opacity when disappearance prevented verification. These behaviors often occurred immediately following moments of resistance, boundary-setting, or requests for verification, rendering disappearance itself a meaningful response rather than a neutral lapse in communication. Mona described how disappearance followed directly after asserting a boundary:
\begin{quote}
\textit{“I asked him to send me a photo; otherwise, I would end the relationship. The moment I made that request, he vanished—for 19 days!”} (Mona)
\end{quote}

Others reported similar experiences of sudden silence following periods of intense communication. Hamid reflected on how this abrupt shift disrupted his assumptions about the relationship:
\begin{quote}
\textit{“After two months in a relationship, she just disappeared, which made me skeptical.”} (Hamid)
\end{quote}

For some participants—Tiam, Amin, and Farah—\textbf{\textit{ghosting}} only became legible as a scam indicator retrospectively, after the perpetrator had already achieved their objective. As Tiam explained:
\begin{quote}
\textit{“While we were in a relationship? No. But when she disappeared, I realized I was a victim.”} (Tiam)
\end{quote}

These accounts illustrate that \textbf{\textit{ghosting}} did not always immediately trigger suspicion. Instead, disappearance often generated confusion, uncertainty, and emotional distress, with its significance as a deception cue becoming apparent only through retrospective sense-making following harm. In this way, \textit{\textbf{ghosting}} functioned less as an immediate warning sign than as a delayed signal whose meaning crystallized over time.






\subsubsection{"Against Epistemic Norms of Romantic Love"}
Another suspicion cue emerged when perpetrators’ behaviors violated participants’ shared expectations about how romantic partners should think, feel, and act—what participants implicitly treated as \textbf{\textit{epistemic norms}} of romantic love. These norms concern what a romantic partner is expected to know, care about, and be responsive to within an intimate relationship, shaping judgments about sincerity and authenticity.

One prominent violation involved the persistent avoidance of face-to-face meetings. Participants consistently described repeated postponements, canceled plans, and excuses that prevented in-person encounters. This pattern was especially salient in cases where scammers’ primary objective was financial rather than sexual exploitation. When perpetrators were not seeking physical intimacy, they appeared to deliberately stall offline escalation while maintaining emotional engagement through digital communication. From participants’ perspectives, this behavior conflicted with core assumptions about romantic relationships, particularly the belief that genuine affection entails a desire for physical presence and mutual visibility. As Melina explained:
\begin{quote}
\textit{“I asked him to have a date in our city, but every time I asked, he made an excuse, and he never came. Lovers should want to see each other, shouldn’t they?”} (Melina)
\end{quote}

In other cases, suspicion arose when perpetrators responded with indifference or emotional detachment to participants’ expressions of fear, stress, or vulnerability. Such reactions violated expectations that romantic partners should demonstrate care, empathy, and emotional responsiveness. Rahil described this breach:
\begin{quote}
\textit{“I could not find any reason for my boyfriend’s indifferent reaction […] he acted like it was not important to him how much pressure I had received.”} (Rahil)
\end{quote}

Across these accounts, suspicion did not stem from a single action but from perceived misalignments between perpetrators’ behaviors and participants’ \textbf{\textit{epistemic expectations}} of romantic intimacy. When actions failed to conform to these norms—whether through avoiding embodied encounters or dismissing emotional vulnerability—participants began to question the authenticity of the relationship and the intentions underlying it.

\subsubsection{Avoiding Social Connections}
Another major suspicion cue involved \textit{\textbf{avoiding social connections}}, whereby perpetrators actively restricted victims’ access to their social networks, concealed their own digital presence, or denied having social media accounts altogether. This form of social isolation can be understood as a strategy of identity opacity, limiting opportunities for external verification. Participants interpreted these behaviors as deliberate attempts to limit opportunities for verification and to prevent exposure to inconsistencies, competing narratives, or other potential victims. In some cases, scammers sought to control victims’ online interactions by discouraging or prohibiting contact with others in their social circle. Shadi described how such control prompted suspicion:
\begin{quote}
\textit{“I suspected something was wrong when he asked me to unfollow all his friends on Instagram. His excuse? He didn’t want me interacting with other guys. [...] But they were his own friends! Keeping me away from them meant he was hiding me—not acceptable!”} (Shadi)
\end{quote}

Other participants encountered avoidance through the concealment of the perpetrator’s own online presence. Mahi became suspicious when an inadvertent mistake revealed a contradiction:
\begin{quote}
\textit{“He told me he wasn’t on Instagram. But one day, when he sent me a screenshot of his phone for something else, I saw the Instagram app right there! That’s when I knew—he had been lying all along.”} (Mahi)
\end{quote}

Across these accounts, \textbf{\textit{avoiding social connections}} functioned as a form of relational isolation that constrained victims’ ability to corroborate identities, observe social behavior, or seek informal validation. Rather than being interpreted simply as privacy preferences, such practices raised suspicion when they conflicted with participants’ expectations that romantic partners should be socially visible and accountable—especially in relationships already reliant on digital mediation. In this way, social avoidance emerged as a salient cue through which participants inferred deceptive intent and began reassessing the authenticity of the relationship.



 \subsubsection{"Sexual Requests: Moral Boundary Violations and Accelerated Suspicion"}
\textbf{\textit{Sexual requests}} emerged as one of the most salient and destabilizing suspicion cues in participants’ accounts. Within the Iranian context, where sexual intimacy outside marriage is legally restricted and socially stigmatized, such requests constituted a direct violation of moral, social, and relational boundaries. Rather than being interpreted as flirtation or normative escalation, sexual demands were widely experienced as signals of deceptive or exploitative intent.

Nearly all female participants reported being subjected to \textbf{\textit{sexual requests}}, including pressure to engage in sexting, send explicit images, or participate in sexual video calls. These demands were often framed by perpetrators as expressions of intimacy or trust-building, yet participants interpreted them as deeply incongruent with expectations of romantic commitment. As Gelare explained:
\begin{quote}
\textit{“When he asked me to fulfill his sexual needs, I suspected.”} (Gelare)
\end{quote}

In cases of what participants described as “body scams,” \textbf{\textit{sexual requests}} functioned not only as deception cues but as mechanisms of coercion. Perpetrators leveraged promises of marriage or emotional attachment to push victims toward compromising situations, exploiting the high personal and social costs associated with sexual exposure. Unlike financial scams, where deception could unfold gradually, these requests often collapsed ambiguity and triggered immediate fear, distress, and reassessment of the relationship.

Suspicion was further intensified when perpetrators insisted on meeting in private spaces. In Iranian dating norms, proposing first meetings in a private home—rather than a public setting—was perceived as highly inappropriate and dangerous. Tina described how such a request prompted fear and mobilized external support:
\begin{quote}
\textit{“He told me I should go to his place and have sex with him! [...] I was very scared, so I asked my sister to help me.”} (Tina)
\end{quote}

Similarly, Rahil interpreted the request as evidence that the relationship was never intended as romantic:
\begin{quote}
\textit{“He said I should go to his home and have sex with him. It was not possible for me. I was terrified. I understood that he did not love me—only sex mattered to him.”} (Rahil)
\end{quote}

Across these accounts, \textbf{\textit{sexual requests}} operated as a decisive boundary violation that transformed uncertainty into recognition. In a context where sexual transgression carries severe social and personal consequences, such demands were not merely suspicious but fundamentally incompatible with participants’ epistemic and moral expectations of romantic love. As a result, sexual requests frequently accelerated scam recognition while simultaneously heightening fear, vulnerability, and the need for protective action.


\subsubsection{Avoiding Authentication: Refused to Meet, Instant Call and Selfie Sending}
\textbf{\textit{Avoiding authentication }}emerged as a salient suspicion cue when perpetrators systematically resisted any form of real-time or embodied verification, including live video calls, selfies, voice calls, or in-person meetings. While participants initially tolerated some delays or refusals as situational or circumstantial, persistent avoidance across multiple modes of authentication became increasingly difficult to reconcile with claims of genuine intimacy. Unlike broader relational norm violations, this category directly concerns the practical impossibility of verifying identity. Many participants described growing skeptical when requests for spontaneous images or live interaction were repeatedly deflected. The absence of real-time visual or auditory confirmation undermined perpetrators’ credibility, as Zoya explained:
\begin{quote}
\textit{“Whenever I asked for a selfie, he never sent one. He also never answered my phone calls.”} (Zoya)
\end{quote}

Beyond digital interaction, continued resistance to meeting in person—particularly after prolonged emotional engagement—further intensified suspicion. Participants noted that repeated postponements appeared patterned rather than coincidental, prompting a reassessment of earlier justifications. Arman reflected:
\begin{quote}
\textit{“She made a lot of excuses. One time she said she needed to study. Another time, she blamed her controlling family. Every single time, she had a reason to avoid meeting.”} (Arman)
\end{quote}

Across accounts, participants distinguished between ordinary hesitation in online dating and systematic \textbf{\textit{avoidance of authentication}}. Whereas delayed meetings or limited availability could be interpreted as caution or situational constraint, sustained resistance to verification across multiple modalities signaled a practical impossibility of confirming the relationship’s authenticity. Unlike violations of \textbf{\textit{epistemic norms}} of romantic love, which generated doubts about sincerity or care, the absence of authentication—across images, calls, and physical encounters—confronted participants with the inability to verify identity, becoming a defining indicator of deception.


\subsection{Red Flags Treated as Provisional: How Suspicion Was Deferred} \label{4.2}
While the cues described in Section \ref{4.1} often generated discomfort or unease, they did not consistently lead to immediate recognition of deception. Instead, many participants interpreted early warning signs as provisional, ambiguous, or explainable, actively rationalizing them in ways that preserved the possibility of a genuine romantic relationship. This section examines how suspicion was deferred—or failed to emerge at all—through emotional investment, epistemic norms of romantic love, and socio-cultural constraints, even in the presence of behaviors that were later recognized as deceptive.

Across accounts, participants described an initial sense that “something felt off” without labeling the interaction as fraudulent. Rather than prompting verification or disengagement, early doubts were frequently neutralized through interpretations that reframed risk as romance, commitment, or necessity. Three recurring mechanisms contributed to this suspension of suspicion: exaggeration was reinterpreted as affection, epistemic norms of love positioned trust as a moral obligation, and societal barriers normalized risky arrangements.

\subsubsection{Exaggeration Reframed as Love}
Although exaggerated affection and emotional intensity were later recognized as suspicion cues, several participants initially interpreted such behaviors as evidence of care, devotion, or genuine romantic interest. Rather than triggering doubt, sustained attention and intimacy often reassured participants of the relationship’s authenticity, preventing suspicion from forming.

Bita’s account illustrates how exaggeration functioned to preclude suspicion altogether. When asked whether she had ever suspected deception, she responded:
\begin{quote}
\textit{“Never! He gave a lot of attention to me. A lot of love. […] We were chatting for around one year almost every night till between 4 and 5 a.m.! Still I do not have any idea how it was possible?”} (Bita)
\end{quote}

Similar rationalizations appeared across other interviews, with participants describing perpetrators as \textit{“very kind,”} \textit{“so polite,”} or emotionally available through constant communication. In these cases, emotional intensity masked implausibility, allowing exaggerated behavior to be read as romantic commitment rather than manipulation. Exaggeration thus functioned not only as a potential red flag, but as a mechanism that delayed or prevented the emergence of suspicion.

\subsubsection{Societal Barriers and the Normalization of Risk}
Socio-cultural constraints further contributed to the deferral of suspicion by making risky arrangements appear necessary or unavoidable. In Iran, public dating is often stigmatized or impractical, particularly in smaller or more conservative regions. As a result, avoiding face-to-face meetings was sometimes considered normal behavior in these contexts. As Rahil explained:

\begin{quote}
\textit{"In our city, if I were seen with a person of the opposite sex in public, even a male neighbor could beat me in the street, and my family would accept it because of ‘family honor.’ He knew this, and he suggested that we either not meet or meet at his house. For me, avoiding meeting in person seemed like normal behavior."}(Rahil)
\end{quote}

Moreover, private or semi-private meetings—while potentially dangerous—were frequently normalized as the only viable option for pursuing a relationship. Participants from smaller cities described accepting invitations to meet in offices, private homes, or secluded locations as a compromise between safety and secrecy. Feri explained:
\begin{quote}
\textit{“My goal was marriage, and I had to meet him. I accepted his invitation to meet at his company. It felt safer than a private house, but also not too public where I could be seen and judged. […] I had no reason to be skeptical.”} (Feri)
\end{quote}

Beyond meeting arrangements, promises of marriage played a central role in suppressing doubt. In contexts where premarital relationships are stigmatized, the prospect of marriage provided moral justification for enduring discomfort and risk. Participants described overlooking inconsistencies and escalating demands because the relationship was framed as a legitimate pathway to marriage. In this way, societal barriers did not merely shape behavior; they actively structured how risk was interpreted, rendering early warning signs less salient and delaying recognition of deception.


\subsection{Recognition Without Exit: Love, Dependence, and the Gamble} \label{4.3}

Whereas Section 4.2 describes how suspicion was deferred or neutralized, this section focuses on cases in which participants recognized deception, harm, or inconsistency but nonetheless remained in the relationship. Even after acknowledging warning signs—or explicitly identifying the interaction as deceptive—some participants continued engagement due to emotional dependence, fear of loss, and structural vulnerability. Recognition, in these cases, did not translate into disengagement.

\subsubsection{Emotional Dependence and Fear of Abandonment}
Several participants described recognizing suspicious behavior yet choosing not to act due to deep emotional attachment. Love and dependence made disengagement feel more threatening than continued uncertainty. Mahi reflected:
\begin{quote}
\textit{“I could feel there was something wrong! But because of my great dependence on him and love, I ignored all of the signs.”} (Mahi)
\end{quote}

In other cases, fear of abandonment was explicitly weaponized. Arman explained how threats to end the relationship silenced his doubts:
\begin{quote}
\textit{“There was something suspicious. Whenever I questioned her, she threatened that she would end the relationship. Since I loved her a lot, I stopped taking action on my suspicions.”} (Arman)
\end{quote}

Here, recognition of risk coexisted with emotional constraint, producing inaction rather than exit.

\subsubsection{The Gamble of Marriage and Structural Vulnerability}
For some participants, remaining in the relationship was framed as a calculated gamble. Emotional investment, combined with the promise of marriage, led victims to tolerate escalating harm in the hope that the relationship would ultimately “pay off.” This was especially pronounced in contexts where leaving a relationship carried significant social consequences.

Shiva described how the prospect of marriage kept her engaged despite recognizing exploitation:
\begin{quote}
\textit{“The chance for marriage stopped me from cutting ties—even with a sexual scammer. I stayed patient, trying to convince him to marry me. I thought I had no choice.”} (Shiva)
\end{quote}

In regions where virginity remains a prerequisite for marriage, the stakes were even higher. Roya explained:
\begin{quote}
\textit{“I thought if he left me, I had no chance for marriage. In our region, nobody agrees to marry a non-virgin girl. […] So I accepted everything he asked, even the unusual things.”} (Roya)
\end{quote}

In these cases, continued engagement reflected not naivety but constrained choice under structural and moral pressure.

\subsubsection{Power Asymmetry and Coercion}
In the most severe cases, recognition of harm occurred alongside ongoing coercion and abuse. Sama described realizing she had been exploited while remaining emotionally bound:
\begin{quote}
\textit{“He had sex with me in his car by force. […] I understood that he wanted me just for sex. But I loved him. After that, I still continued communicating with him.”} (Sama)
\end{quote}

These accounts highlight stark power imbalances between emotionally vulnerable victims and manipulative perpetrators. Recognition alone was insufficient to enable exit; emotional attachment, fear, and social constraint combined to trap participants in harmful dynamics.


\subsection{Investigation Strategies: User-Led Attempts to Establish Truth} \label{4.4}
When suspicions emerged, participants rarely disengaged immediately. Instead, many undertook active efforts to establish the truth of the relationship. These investigative practices were shaped not only by the affordances of digital platforms, but also by socio-cultural constraints that limited access to offline verification, discouraged confrontation, and made disclosure risky. In contexts where public dating, direct questioning, or involving authorities could carry social or legal consequences, investigation often unfolded cautiously, indirectly, and over extended periods of time.

\subsubsection{Searching Across Digital and Physical Contexts}
One of the most common investigative strategies involved searching for information across digital platforms and, in some cases, physical environments. Participants conducted internet searches using names, photos, phone numbers, and workplace details, often cross-checking scammers’ narratives against information found on social media or search engines. For participants who had never met the perpetrator in person, online searching was often the only available means of verification. Shadi described using Google and Instagram to uncover inconsistencies in her scammer’s age and identity, while Mahi identified her offender’s real name through Instagram. Zoya discovered a fundamental deception when a photo sent by the scammer revealed a tagged Instagram account:
\begin{quote}
\textit{“By searching online. When I noticed abnormal behaviors, I wanted to find out the truth. One time, when he sent me a picture and said it was his photo, I suddenly saw an Instagram ID tagged on that image. By messaging that account, I found the truth.”} (Zoya)
\end{quote}

Other participants extended their searches beyond mainstream platforms. Sama searched Telegram groups connected to her scammer’s village, while Betty contacted acquaintances to corroborate information. In rare cases, participants who had met their scammer in person conducted physical searches. Hamid described searching through the perpetrator’s belongings to uncover identifying documents. These accounts illustrate how investigation often required creativity and persistence, especially when access to direct verification was restricted. In a context where face-to-face verification was often delayed, avoided, or socially constrained, online searching became a critical substitute for offline authentication, allowing participants to investigate without escalating risk or drawing attention to the relationship.


\subsubsection{Questioning and Testing the Scammer}
Another investigative approach involved directly or indirectly questioning the perpetrator. Participants described repeatedly requesting face-to-face meetings, identification documents, or verifiable social media profiles, particularly when authentication had been persistently avoided. Refusals or evasive responses frequently intensified suspicion, though they did not always result in immediate disengagement.

Some participants adopted indirect tactics to test consistency, such as fishing for information through fabricated scenarios or alternate accounts. Hamid described creating a false narrative to provoke a reaction:
\begin{quote}
\textit{“I pretended I had a friend in the Security and Information System Organization. She became scared and thought I knew everything about her.”} (Hamid)
\end{quote}

Direct confrontation—explicitly accusing the scammer of deception—was typically a last resort. Participants described delaying confrontation due to fear of losing the relationship, concern about retaliation, uncertainty about their conclusions, and the risk of social exposure should the interaction escalate or become visible to others. As a result, questioning often functioned less as a decisive intervention and more as a gradual process of sense-making under emotional and socio-cultural constraint.

\subsubsection{Requests for Proof and the Production of False Evidence}
Participants also sought verification through requests for proof documents, such as identity cards, tickets, or evidence of travel, as well as real-time selfies, video calls, or in-person meetings. While these requests were intended to resolve uncertainty, they often resulted in the production of fabricated evidence.

Shadi described receiving a falsified identification card that had been digitally altered, while Betty recounted being repeatedly shown fake train tickets as proof of an upcoming meeting:
\begin{quote}
\textit{“Many times he sent me photos of tickets from his city to mine, but all were fake. One time he even admitted it. He said he only reserved the ticket to show me and then canceled it.”} (Betty)
\end{quote}

Requests for proof carried heightened stakes in this context, as failed verification could not easily be followed by public accountability or institutional recourse, increasing the cost of being wrong. Even when participants insisted on face-to-face meetings, scammers frequently agreed in principle while repeatedly canceling at the last moment, prolonging the interaction without exposure. Requests for selfies or video calls were similarly circumvented through excuses or manipulated images. These experiences demonstrate how scammers actively adapted to victims’ investigative efforts, producing convincing but deceptive artifacts that further delayed recognition.

\subsubsection{The Role of Third Parties in Detection and Recovery}
Family members, friends, colleagues, and professionals played a critical role in many participants’ investigative processes. Third parties often identified inconsistencies or risks earlier than victims themselves and, in some cases, conducted independent investigations.

Several participants described family members becoming suspicious once relationships were disclosed. Friends and relatives frequently searched for information, warned participants, or encouraged disengagement. Shadi recalled how her psychologist uncovered extensive information about her scammer, although she initially struggled to accept it:
\begin{quote}
\textit{“In the next meeting with my psychologist, she told me she had found a lot of information about my boyfriend—his real photo, age, job, education, and other details.”} (Shadi)
\end{quote}

In other cases, third parties directly intervened. Tiam credited her father with uncovering the truth, while Melina’s sister traced the ownership of a SIM card to expose the scammer’s identity. Shiva described how a sociology teacher’s warnings prevented further escalation:
\begin{quote}
\textit{“I’m grateful that my teacher stopped me from taking things further. Otherwise, I could have been killed.”} (Shiva)
\end{quote}

Roya’s experience illustrates how third-party support could also enable recovery. With encouragement from her dormitory friends, she ended the relationship, consulted a lawyer, and—by reporting the incident as financial fraud rather than a romantic relationship—was able to avoid potential legal consequences and ultimately recover her stolen money. Across accounts, third parties often recognized suspicious behavior earlier than victims themselves and actively intervened by searching for information, issuing warnings, or encouraging disengagement. These interventions sometimes prevented further harm or re-victimization, though in other cases victims initially resisted, caught between emotional investment and the unsettling implications of the evidence presented. In a context where victims were often reluctant to involve formal authorities, trusted third parties—such as family members, teachers, or close friends—functioned as crucial epistemic intermediaries, helping to counteract emotional dependence and providing both informational resources and moral permission to reassess the relationship.

All in all, these strategies demonstrate that victims were not passive recipients of deception, but engaged in sustained and often creative investigative work under conditions of emotional attachment, uncertainty, and constraint.

\section{Discussion}
This study examined how victims of online dating romance scams in Iran retrospectively made sense of deception and responded to emerging suspicion. While prior research in fraud victimology, deception detection, and social engineering \cite{cross2022suspect, cross2023knew, lacey2024cyberpsychology, alizadeh2023catch, kopp2017your} with predominantly Western samples \cite{wang2026scoping, cross2018marginalized} has shown that scam detection often unfolds gradually rather than through isolated red flags \cite{whitty2015anatomy, cross2018denying, cross2023knew, kopp2016online, kopp2015role}, our findings elucidate how local cultural dynamics uniquely shape the incremental and gradual suspicion-turned-detection process.

In section 5.1 we discuss the implications of local cultural dynamics on how online dating romance scam suspicion and detection unfolds, specifically to illustrate how similarly gradual processes of detection across Western and non-Western cultures can unfold under very different, culturally-specific forces. In our particular non-Western context of Iran, socio-cultural norms around premarital relationships, legal risks associated with disclosure, and expectations surrounding marriage fundamentally reconfigure how suspicion is interpreted and acted upon. Rather than simply delaying detection, these conditions actively reshape the detection process itself: participants often avoided direct verification (e.g., face-to-face meetings), tolerated uncertainty in pursuit of socially valued outcomes such as marriage (as captured in “The Gamble of Marriage and Structural Vulnerability”), and relied on trusted others to interpret ambiguity. This extends prior models of fraud detection by showing that the trajectory from suspicion to detection is not universal, but contingent on local cultural norms and platform choices.

In section 5.2 we discuss implications for technology design to mitigate and intervene in online dating romance scams in Iran. We center the Iranian context to first argue why dating app-specific design interventions would be ineffective in Iran given the distributed and distinctly\textit{ non}-dating platforms used by romance scammers in this context. We then argue for third party platform approaches that support situated, emotionally constrained, and culturally specific detection work by augmenting distributed support from family, friends, and local strangers for knowledge sharing about detecting scams in culturally specific contexts and for disentangling from emotional dependence to scammers.

\subsection{Culture as an Interpretive and Action Constraint on Online Dating Romance Scam Suspicion and Detection}

By foregrounding victims’ narratives, this study advances understanding of how romance scams are recognized and acted upon within high-risk, socially constrained (and makeshift) dating environments in a non-Western culture.  Rather than conceptualizing detection as the moment when a red flag is noticed, our findings demonstrate that detection unfolds through iterative sense-making and investigative labor, often  under conditions where confrontation, disclosure, or formal reporting carry significant personal risk. 

An iterative, gradual process of scam investigation and discovery has been found in prior work \cite{lacey2015taking, lacey2024cyberpsychology, kopp2015role, kopp2016online, cross2016they, rege2009s}, however a central contribution of this study lies in demonstrating how local (non-Western) cultural and legal conditions shape both what is recognized as suspicious and what actions are considered possible or safe. In other words, the plodding nature of romance scam detection in our study was not simply a matter of personal or internal qualities like emotional attachment to the scammer or  cognitive biases as other work would suggest, but of factors external to the victim imposed through social norms of the immediate culture.

\subsubsection{The Implications of Iranian Cultural Dynamics on Victims' Suspicions and Detections}

Extending prior work on deception detection \cite{whitty2012online, sorell2019online, atkins2021epistemic}, our findings show that suspicion is frequently deferred or neutralized through emotional and normative mechanisms. Yet in Iran, where premarital romantic relationships are legally restricted and socially stigmatized, behaviors such as sexual requests, pressure for secrecy, or avoidance of public meetings carried different meanings and consequences than they might in more permissive contexts. For many participants, sexual requests functioned as particularly salient suspicion cues; not only because of personal discomfort, but because such requests violated deeply held moral and social norms. At the same time, these constraints shaped victims’ responses to suspicion. Participants described avoiding direct confrontation, delaying disclosure, or reframing their experiences (e.g., reporting financial fraud rather than romantic involvement) to mitigate potential legal or social repercussions. In a more literal or practical sense, cultural norms were also deeply intertwined with the affordances of social media platforms given the banning of dating apps in Iran and thus the necessary repurposing of general social media platforms like Instagram and Telegram into makeshift dating platforms. As a result, identity is fragmented across services, interactions move between public and private channels, and opportunities for verification are unevenly distributed.

Building on this, our findings further suggest that “vulnerability” in intimate interactions is culturally constructed rather than universal. In our context, premarital relationships are often implicitly oriented toward marriage as a socially legitimate outcome. As a result, promises of marriage can function as a powerful mechanism of manipulation, with some participants tolerating uncertainty or taking significant risks in the hope of achieving this outcome, as described in our finding of “The Gamble of Marriage and Structural Vulnerability.” Such dynamics may not carry the same vulnerability in contexts where romantic relationships are less tightly coupled to marriage \cite{amirkhani2025society}.

Similarly, prior work often frames face-to-face meetings as a key strategy for verifying identity and reducing deception \cite{jimoh2018love}. However, in our context, such interactions may themselves introduce risk, leading users to avoid them or accept private settings, as described in our findings "Societal Barriers and the Normalization of Risk". Scammers may exploit these conditions by repeatedly avoiding in-person meetings or by proposing meetings in secluded environments where social visibility is reduced. These examples highlight that common indicators of vulnerability cannot be assumed to translate across contexts. Designing for online dating therefore requires attention to how cultural norms shape both user behavior and the strategies of deception.

\subsubsection{Romance Scam Detection as Epistemic Labor}
One of the most pronounced differences in romance scam detection strategies in our Iranian sample compared to prior research is coordination with third parties for scam confirmation and recovery. Prior research of online dating romance scams in Australia has also found that some victims seek "assistance from a third party" after initial suspicion \cite{cross2022suspect}, however this strategy was relatively rare (used in only 5\% of cases \cite{cross2023knew}). In our study it was much more common, and critical for many of our participants during their investigate processes, in part because law enforcement and other formal authority figures were not safe sources for support. Another key difference was which types of third parties were consulted for assistance. Prior work identified friends and family as typical sources for assistance and verification of suspicions. Our Iranian sample also found friends and family to be common for this role, but the types of stakeholders consulted with were much more diverse and wide-ranging, including neighbors in one's dormitory, a psychologist, a lawyer, and other individuals that would not fit cleanly in a "friend" or "family" category.

The critical, sustained, and intricate nature of third party support roles in our study portray scam detection in Iran as a form of collaborative epistemic labor \cite{hardwig1985epistemic, goldberg2011division}; a term coined in philosophy and particularly social epistemology that describes cognitive tasks that are collaborative and distributed across multiple people. Our findings show such third parties being deeply involved in romance scam investigations, such as by doing independent information searches on their own, some of which were highly technical and nuanced like tracing ownership of a SIM card. In other cases the epistemic labor was more emotional in nature, serving as a stand-in for the emotional dependence victims had on their scammers so it would be easier to discontinue contact, and in other cases by playing more forceful roles in stopping interaction between victim and scammer. 

What distinguishes these activities as epistemic labor from other forms of third party involvement in prior work was in the way the labor was carried out iteratively, under \textit{shared} emotional and time investment. These strategies were rarely linear or decisive; instead, they unfolded incrementally as participants weighed competing interpretations and assessed the risks of being wrong.

Friends, family members, teachers, and mental health professionals were frequently essential epistemic intermediaries to ending scams, identifying inconsistencies earlier than victims themselves and providing alternative interpretations that challenged emotionally invested narratives. In a setting where involving formal authorities (especially law enforcement personnel) could be risky, trusted third parties played a critical role in enabling recognition, exit, and, in some cases, recovery. In this sense, our contribution is not to suggest that victims involve third parties in scam investigations, but to articulate the depth and critical role that shared investigative practices play in romance scams in our study's Iranian context.

\subsection{Design Implications for Supporting Online Dating Romance Scam Detection in Iran}

\subsubsection{An Argument Against Dating App-Specific Design Interventions}

There are myriad design interventions for dating apps that could potentially preempt romance scams through, say, identity verification and AI-driven pattern recognition of scammers, or improved user reporting features. Yet it is important to keep in mind that dating apps are already banned in our study context of Iran. Dating app-specific features would therefore be inaccessible to Iranians by default. Scammers in our study were instead finding victims through more general social media platforms that were repurposed for dating such as Instagram and Telegram. But implementation of scam-specific intervention features as mentioned above would be impeded by the multi-faceted uses for general social media platforms compared to the relatively singular pursuit of romance on dating apps.

Perhaps a more practical barrier to implementation of any platform-specific scam mitigation features is cooperation of each social platform. Romance scam experiences found in our study were often distributed across multiple social platforms, which would require buy-in and cooperation from multiple companies to implement cross-platform scam mitigation features. At least some of these companies like Telegram have taken a vocal stance against user moderation, which would make such proposals an immediate non-starter.

Ultimately, design implications \textit{for} the platforms that romance scams occur through in Iran are almost certainly futile. This is a markedly different conclusion than prior work with victims in Western cultures where scam victimization is taken seriously, as a criminal offense in many cases, and thus has an inherent sense of urgency for unified intervention from legal and law enforcement entities \cite{cross2020romance, USA_PROTECT_2003, UK_SexualOffences_2003, Australia_OnlineSafety_2021,EU_Directive_2011}. We instead recognize that any meaningful intervention into online dating romance scams in Iran (and potentially other non-Western cultures) would need to occur through third party applications designed specifically for scam mitigation.

\subsubsection{Third-Party–Mediated Risk Recognition as Compensation for Platform Limitations}

A key finding is that deception recognition is often mediated by trusted third parties rather than through platform mechanisms. Participants described situations and sought interpretations from friends, siblings, and trusted others, who acted as epistemic intermediaries in making sense of ambiguous interactions and identifying patterns of deception. This reliance highlights the limitations of platform mechanisms such as reporting and verification, which are ill-suited to addressing emotionally embedded and ambiguous forms of deception. As a result, detection is distributed across informal support networks that provide both interpretive and emotional support.

These practices function as a compensatory layer for gaps in platform design, suggesting that detection should be understood—and designed for—as a distributed, socially grounded process rather than an individual or platform-driven task. This points to opportunities to support such practices through tools that enable selective or anonymized sharing of interaction excerpts (e.g., messages) or lightweight mechanisms for seeking second opinions from trusted contacts. However, such interventions should avoid over-formalizing what is currently effective precisely because it remains low-risk, informal, and socially embedded, particularly in sensitive socio-cultural contexts.

Overall, these implications shift the focus from platform-centric prevention toward resilience-oriented, user-driven, and distributed safety infrastructures. By grounding design in how users already navigate suspicion under emotional and socio-cultural constraint, this work contributes actionable HCI directions for supporting detection as an ongoing, situated process.

\section{Conclusion}
Romantic relationships mediated through digital platforms pose fundamental challenges for verifying identity, intention, and authenticity \cite{cross2022suspect, cross2023knew}. In the absence of embodied cues, users must rely on textual interaction, selectively shared media, and fragile forms of trust—conditions that romance scammers deliberately exploit. This study examined how victims of online dating romance scams retrospectively experienced, interpreted, and responded to deception, focusing on how suspicion emerged and when, how, or whether it became actionable detection.

Drawing on in-depth interviews with Iranian online daters, we showed that detection is not a singular moment of realization but a gradual, interpretive process shaped by emotional investment, epistemic norms of romantic love, and socio-cultural and legal constraints. Victims often encountered multiple warning signs without immediately labeling the interaction as deceptive, treating suspicion as provisional and negotiating its meaning over time. When doubt persisted, participants engaged in sustained user-led investigative and a form of collaborative epistemic labor—searching across platforms, testing the scammer, requesting proof, and mobilizing trusted third parties—often under conditions where confrontation, disclosure, or formal reporting carried significant personal risk.

By situating these processes within the Iranian context, this study demonstrates how cultural norms and legal restrictions shape both what behaviors are recognized as suspicious and which investigative actions are considered possible or safe, that is offering a culturally grounded, victim-centered account of how deception is recognized, resisted, and sometimes endured. Importantly, our findings challenge deficit-based portrayals of victims by foregrounding their agency, creativity, and resilience, even in highly constrained environments.

In conclusion, this work underscores the need for prevention efforts, educational initiatives, and platform designs that move beyond awareness messaging alone. Particularly in contexts where traditional dating apps are absent, legal recourse is limited, or disclosure carries stigma, interventions should support low-risk verification practices, distributed social support, and third-party–mediated detection work may better equip individuals to detect and respond to romance scams across diverse cultural settings. Future research should continue to examine detection as an interpretive and relational process, attending to how local norms, technologies, and power structures shape both vulnerability and resistance to online deception.

\begin{acks}
The authors would like to thank all study participants for their time and valuable participation. We also thank the anonymous reviewers for their constructive comments. This work was supported by the German Federal Ministry of Education and Research (BMBF) as part of the AntiScam project (Grant No. 16KIS2214). It was also partially supported by the U.S.National Science Foundation under Grant Nos. 2211896, 2401775, and 2339431. 
\end{acks}

\bibliographystyle{ACM-Reference-Format}
\bibliography{references}

\appendix
\section{Data Availability Statement}

The data that support the findings of this study are not publicly available due to confidentiality agreements with participants and the sensitive nature of the collected information. We only show some demographic information in this table:
\begin{table*}
  \caption{Anonymized identifiers Participant Overview}
  \label{tab:commands}
  \begin{tabular}{p{3cm} p{2.5cm} p{2.5cm} p{ 2.5 cm}}
    \toprule
    Pseudonyms 

(Gender)& Age

(Incident|Interview)& Scam Duration& Interview\\
    \midrule
    1. Mona (F)  & 18 | 20 & 2 months &  T: Chat \\
    2. Shadi (F)  & 16 | 20 & ~4 years &  T: Voice \\
    3. Farah (F)  & 19 | 25 & 1 months &  I: Chat \\
    4. Hamid (M)  & - | 28 & 5 years & W: Voice \\
    5. Mina (F)  & 17 | 19 & 2 months &  W: Call \\
    6. Mahi (F)  & 16 | 21 & 5 months &   T: Chat \\
    7. Zoya (F)  & 16 | 20 & 3 years & T: Voice \\
    8. Malina (F)  & 17 | 18 & ~1 year & T: Chat \\
    9. Tina (F)  & 14 | 18 & 3 months &  T: Chat \\
    10. Ava (F)  & 15 | 19 & 6 months & T: Chat \\
    11. Sama (F)  & 31 | 31 & 2 months & T: Chat \\
    12. Gelare  (F)  & 31 | 34 & ~3 years & T: Chat \\
    13. Betty  (F)  & 19 | 22 & 2 months & I: Chat \\
    14. Tiam  (M)  & 19 | 21 & 6 months & T: Chat \\
    15. Amin  (M)  & 17 | 19 & 2 months & T: Chat \\
    16. Feri  (F)  & 36 | 39 & ~1 months & T: Chat \\
    17. Bita  (F)  & 17 | 25 & ~1 year & T: Chat \\
    18. Gilda  (F)  & 22 | - & ~1 year & T: Chat \\
    19. Rahil  (F)  & 16 | 26 & 2 years & T: Chat \\
    20. Sara  (F)  & 28 | 28 & 2 months & T: Chat \\
    21. Shiva (F)  & 17 | 19 & 3 months & T: Call \\
    22. Hiva (F)  & 26 | 27 & 1 Week & T: Video call \\
    23. Roya (F)  & 21 | 25 & ~2 years & T: Call \\
    24. Arman (M)  & 21 | 23 & ~2 months & T:Chat \\
    \bottomrule
  \end{tabular}
\end{table*}

\end{document}